\documentclass[12pt]{article}

\usepackage{sbc-template}
\usepackage{graphicx,url}
\usepackage[utf8]{inputenc}
\usepackage[english]{babel}
\usepackage{float}

\title{Usable but Conventional: An Empirical Study on the UX of AI-Generated Interface Prototypes}

\author{Karoline Romero\inst{1}, Igor Wiese\inst{2}, Renato Balancieiri\inst{1} \\ Gislaine Camila Leal\inst{1},
Guilherme Guerino\inst{1,3}}

\address{Universidade Estadual de Maringá (UEM)\\
  Maringá, Paraná
\nextinstitute
  Universidade Tecnológica Federal do Paraná (UTFPR)\\
  Campo Mourão, Paraná
\nextinstitute
  Universidade Estadual do Paraná (UNESPAR)\\
  Apucarana, Paraná
  \email{\{karolharummy,igor.wiese\}@gmail.com, 
  \{rbalancieri,gclleal\}@uem.br}
  \email{
  guilherme.guerino@ies.unespar.edu.br}
  }

\begin{document} 

\maketitle

\begin{abstract}
This paper investigates User Experience (UX) with prototypes generated by Generative Artificial Intelligence (GenAI) tools. An empirical survey with 92 participants evaluated AI-generated and human-created prototypes without prior identification of authorship. We measured UX using the UEQ-S, covering pragmatic and hedonic dimensions. Results indicate positive evaluations in pragmatic aspects, such as usability and efficiency, and neutral or negative evaluations in hedonic aspects, including originality and innovation. We concluded that GenAI can produce functional interfaces but tends to reinforce visual and structural patterns that affect perceptions of originality.
\end{abstract}
     
\section{Introduction}

The use of Artificial Intelligence (AI) has increased in different fields of design. Graphic designers use AI to expand their knowledge in semiotics, typography, and layout, while User eXperience (UX)/User Interface (UI) designers employ these technologies in the design of interfaces and user experiences on \textit{websites} and mobile applications \cite{10.1145/3610217}. Industrial designers use AI to support aspects related to manufacturing, material use, and ergonomics. Although AI alters the way design tasks are performed, its main impact is to complement and expand human creativity rather than replace it \cite{10.1145/3610217}.

In this sense, the advancement of Generative AI (GenAI) has contributed to the development of creative processes, assisting designers in stages such as empathy, ideation, and prototyping \cite{10382802, ihcgab}, suggesting layouts, identifying patterns, and automating tasks such as information organization \cite{TROCIN2023122724}. However, the use of generative AI technologies raises ethical, technical, and operational concerns related to transparency, accountability, biased outputs, and the quality and representativeness of training data. Moreover, responsibility for harms or misuse generated by AI systems cannot be delegated to machines, reinforcing the importance of continuous human supervision and governance mechanisms throughout the AI lifecycle \cite{pasetti2025}.

Therefore, our paper aims to investigate whether the UX differs between interface prototypes created by humans and those created by GenAI tools. 
We used an empirical evidence-based approach, structured through a survey, following the guidelines of de \cite{Kitchenham2008}. The study used primary data collected directly from participants. UX was evaluated using the User Experience Questionnaire – Short Version (UEQ-S) \cite{schrepp2023ueq}, considering pragmatic and hedonic aspects. Data were processed by calculating averages for each item and aggregating them across pragmatic and hedonic quality dimensions \cite{schrepp2023ueq}. Based on these calculations, we compared the prototypes' results across dimensions and identified trends in evaluation patterns. 

Furthermore, the analysis sought to verify the existence of an association between perceived UX and the authorship of the prototypes, exploring whether the origin, human or AI, significantly influenced the UX perceived by the participants. In this sense, the contributions of this article are: i) an empirical comparison of UX perception between AI-generated prototypes and prototypes designed by humans using UEQ-S; ii) evidence that pragmatic quality dominates UX perception in early-stage prototypes; and iii) identification that AI prototypes tend to reinforce conventional interface patterns, leading to lower hedonic perception. 

This work is organized into four additional sections. Section \ref{sec:trabrel} presents Related Work, with research on the use of Generative Artificial Intelligence in interface design, the AI tools used in our work, and the gap investigated. Section \ref{sec:met} describes the Method, detailing survey planning, prototype generation process, participant profile, and data analysis procedures. Section \ref{sec:result} presents the Results and Discussion, presenting the characterization of the sample, the UX analysis using UEQ-S, the statistical tests applied, and the limitations of the work. Finally, Section \ref{sec:cons} presents the Final Considerations, highlighting the study's contributions and opportunities for future research.

\section{Related Work}
\label{sec:trabrel}

The use of GenAI in interface design has been investigated from different perspectives, ranging from the automation of the creative process to cognitive and organizational impacts on professional practice. For example, the work of Lee (2025) proposes an operational definition of Generative User Interfaces (GenUI) and GenUI design as a distinct design paradigm. The author differentiates GenUI from previous approaches, such as adaptive, customizable, intelligent, or AI-assisted interfaces, identifying five constitutive elements: computational co-creation, expanded exploration of the design space, representational fluidity, contextual adaptation, and synthesis rather than selection. He argues that GenUI represents a new mode of interface creation, marked by collaboration at design time between human agents and machines, and by interaction at usage time with AI-generated variants, configuring a socio-technical process of co-creation \cite{Lee_2025}.

Tholander and Jonsson (2023) investigates how large language models and GenAI systems can support ideation activities in interaction design. The study was conducted through a workshop in which professional and academic designers used GPT-3 to develop ideas from a scenario focused on the experience of families in emergency room waiting rooms. The main result indicates that AI can support the creative process by saving time, quickly mapping the design space, generating initial ideas, and identifying flaws in proposals. The potential to generate complementary materials, such as scenarios and personas, was also recognized. However, participants expressed distrust in AI's ability to autonomously produce innovative, high-quality solutions, attributing this limitation to a lack of in-depth contextual understanding. AI was perceived as more effective at exploring possibilities (going wide) than at deepening solutions (going deep). The authors further discuss how conversational metaphor influences the UX and argue that post-humanist and "more-than-human" perspectives may favor co-creation configurations that better explore the differences between human and computational capabilities \cite{10.1145/3563657.3596014}.

Li et al. (2024) explores the potential impact of GenAI tools on UX Design (UXD) practice from the perspective of UX designers. The study seeks to understand how these professionals perceive the incorporation of GenAI into their workflows, as well as the opportunities and risks envisioned for the future of human-AI collaboration. The results indicate that experienced designers demonstrate confidence in their originality, creativity, and empathic capacity, perceiving GenAI predominantly as an auxiliary tool. They highlight "agency" and "enjoyment" as essentially human dimensions, arguing that ultimate control remains with the designer. However, concerns arise about the impacts on novice designers, including potential skill degradation, professional replacement, and creative burnout. The article also points to implications for responsible collaboration, especially regarding copyright, intellectual property, preservation of human creativity, and AI literacy \cite{Li10.1145/3613904.3642114}.

Wang et al. (2024) investigates how the integration of Artificial Intelligence-Generated Content (AIGC) tools influences social dynamics in UX collaboration and professionals' expectations regarding the future development of these tools and teams. The results indicate two main impacts: on the one hand, the tools contribute to mitigating conflicts by quickly visualizing concepts, offering additional perspectives, and assisting in understanding the workload; on the other hand, they introduce potential conflicts when the generated content is perceived as superior to human content or when there is skepticism about the reliability of the results. The study also points out that the use of AIGC promotes a more exploration and sharing-oriented culture, expanding the roles of some professionals to include instruction and leadership in the use of the tools \cite{10.1145/3643834.3660703}.

The work of Bertao and Joo (2021) explores Brazilian UX/UI professionals' perceptions of incorporating AI technologies into their design practices through a survey. The results indicate that the Brazilian industry is still in an early stage of adoption, with few concrete opportunities to work with intelligent systems and a limited understanding of their potential. The perception of AI as an operational tool that focuses on facilitating process steps and increasing efficiency, especially in automation and data processing activities, predominates. However, participants believe that, in the medium term, the use of AI-based resources will become common in design activities \cite{Bertao20144}.

Despite existing research, a gap persists in the analysis of user perception, particularly in identifying authorship and assessing the influence of AI-generated patterns on UX. In this context, this study proposes investigating how users perceive prototypes generated by AI and humans, analyzing UX and patterns that can lead to confusion or inaccuracies in attributing authorship.

\subsection{Used GenAI tools}

For this research, prototyping tools that incorporate GenAI to generate interfaces from prompts were considered. The tools used were: i) Figma/UX Pilot\footnote{https://www.figma.com/}; ii) Uizard\footnote{https://uizard.io/}; iii) Stitch\footnote{https://stitch.withgoogle.com/}; iv) Lovable\footnote{https://lovable.dev/}; e v) Magic Patterns\footnote{https://www.magicpatterns.com/}. Figma is a collaborative design platform, and in this research, the UX Pilot plugin\footnote{https://uxpilot.ai} was used. It was necessary to select parameters such as design mode, device type, and the number of screens before entering the prompt to generate the prototype. Uizard is a design tool that uses AI to create prototypes, allowing users to choose the device type and specify desired visual characteristics to start the generation. Stitch allows the creation of user interfaces from natural language descriptions or images, enabling direct pasting into Figma and exporting the front-end code \cite{stitch2025}. Lovable allows the creation of complete websites using natural language, with direct prompt insertion. Magic Patterns generates interfaces and visual components using AI and can also be used as a Figma plugin.

\section{Research Method}
\label{sec:met}
This section presents the methodological procedures adopted in this research, structured according to the guidelines in Kitchenham and Pfleeger (2008) for constructing and conducting surveys \cite{Kitchenham2008}. The methodology was organized to cover everything from defining the context for generating prototypes to structuring, applying, and analyzing the survey, also covering the process of creating the artifacts used. The objective of the survey was to understand how users evaluate the UX of prototypes and whether evaluations differ between prototypes created by humans and those generated by AI.

The survey's structuring began with defining the system's context to be prototyped, specifying the application domain, target audience, and essential tasks, and guiding the construction of prompts to maintain coherence across the generated prototypes. As mentioned in the previous section, we selected five tools with GenAI functionalities for desktop prototyping: Figma (UX Pilot plugin), Uizard, Stitch, Lovable, and Magic Patterns. The defined scenario was a desktop system aimed at personalizing activities for students with Autism Spectrum Disorder, focusing on creating individualized educational plans and integrating an LLM-based chat to automatically generate pedagogical proposals. In this sense, a standardized prompt was created in English for use in all tools:``\textit{Desktop system that receives a student profile previously defined by anamnesis and protocols that has a screen with chat integrated into an LLM that generates a proposal for pedagogical activity based on the student's profile}''. 

In addition to the five prototypes created by the GenAI tools (each tool created one prototype), we used another five prototypes for the same context, without the support of GenAI tools. These prototypes were created and refined by a collaborating researcher with experience in UX design. Then, in total, we used ten prototypes\footnote{https://figshare.com/s/65bd0cdfe25590c8f4d6} in the survey, five generated by AI tools and five created by human. The prototypes were interleaved between AI-generated and human-generated ones and presented in a randomized order in the questionnaire to avoid bias. We coded the prototypes by letters A to J and distributed by authorship. Prototypes A, D, E, H, and J were developed by humans, while prototypes B, C, F, G, and I were generated by GenAI tools, as shown in Table \ref{tab:gabarito_ferramenta}. Then, our study used a within-subjects design, in which each participant evaluated all prototypes, i.e., all participants viewed and evaluated the prototypes in the same presentation order, and each participant evaluated all ten prototypes. 

\begin{table}[ht]
\footnotesize
\centering
\caption{Prototype codes, their origin, and the GenAI tool used.}
\begin{tabular}{|c|c|c|}
\hline
\textbf{Prototype} & \textbf{Origin} & \textbf{GenAI tool} \\ \hline
A & Human & -- \\ \hline
B & AI & Stitch \\ \hline
C & AI & Uizard \\ \hline
D & Human & -- \\ \hline
E & Human & -- \\ \hline
F & AI & Figma (UX Pilot) \\ \hline
G & AI & Lovable \\ \hline
H & Human & -- \\ \hline
I & AI & Magic Patterns \\ \hline
J & Human & -- \\ \hline
\end{tabular}
\label{tab:gabarito_ferramenta}
\end{table}


We conducted the survey online via Google Forms, ensuring anonymity and enabling wide dissemination. The questionnaire design aimed to present the prototypes neutrally, without indicating their origin or the tool used to create them. The instrument\footnote{https://forms.gle/qpXAwvYmbgPKUvAr5} included demographic data and an evaluation of the UX of each prototype using the UEQ-S \cite{schrepp2023ueq}. The UEQ-S consists of eight pairs of adjectives that measure pragmatic and hedonic quality, and allows interpretation on a scale of -3 to +3, with values between -0.8 and 0.8 being neutral, above 0.8 positive, and below -0.8 negative \cite{schrepp2023ueq}. Therefore, for each of the ten prototypes provided in the form, each participant completed the UEQ-S to report perceived UX analyzing the artifact. 

The final data collection yielded 92 participants, all students from Computer Science, Informatics, and Electronic Engineering courses. The motivation for selecting students from the computer science and engineering fields as participants is that this profile encompasses both the user aspect and the perspective of future specialists who already possess initial technical knowledge in the area. Participation was voluntary and anonymous, and the research was approved by the Ethics Committee of State University of Paraná (CEP/UNESPAR), registered under CAAE nº 88898725.0.0000.9247.

Data analysis was conducted using the official UEQ-S analysis tool\footnote{https://www.ueq-online.org/}. We inserted responses into the tool, which calculated averages for pragmatic quality, hedonic quality, and the overall average. After calculating the averages, we applied the Shapiro-Wilk test \cite{10.1093/biomet/52.3-4.591} that indicated a violation of normality in the sample, prompting the use of non-parametric tests. Then, we used the Friedman test \cite{Friedman01121937} to verify significant differences between the overall averages, followed by multiple comparisons using the Durbin-Conover method \cite{conover1999}. 

\section{Results}
\label{sec:result}

The results are derived from the application of the methodological process described and include: i) demographic data for sample characterization (N = 92); ii) UX analysis of the prototypes using UEQ-S; and iii) verification of significant differences between the prototypes evaluated.

\subsection{Demographics}

Demographic data characterize the participants who responded to the survey. The majority of respondents (68.5\%, N = 63) were aged 20 to 29 years, while 30.4\% (N = 28) were under 20 years, and 1.1\% (N = 1) were aged 40 to 49 years. Most participants identified as male (82.6\%, N = 76); a smaller group (17.4\%, N = 16) identified as female. Regarding undergraduate courses, nearly all (97.8\%, N = 90) were enrolled in Computer Science, with only 1.1\% (N = 1) each in Electronic Engineering and Informatics. As for year of graduation, 26.1\% (N = 24) were in the 1st year, 19.6\% (N = 18) in the 2nd, 30.4\% (N = 28) in the 3rd, 21.7\% (N = 20) in the 4th, and 2.2\% (N = 2) in the 5th year.

Regarding experience with software development, 39.1\% (N = 36) have never participated in software development projects, 47.8\% (N = 44) have participated in courses, 9.8\% (N = 9) have participated in industry work a few times, and 3.3\% (N = 3) work in the field and consider themselves experienced. Regarding experience in Human-Computer Interaction, 26.1\% (N = 24) have never had contact with the area, 62\% (N = 57) have had occasional contact, and 12\% (N = 11) have daily contact. Regarding GenAI, 22.8\% (N = 21) have never had contact, 32.6\% (N = 30) have used it occasionally, 39.1\% (N = 36) use it daily, and 5.4\% (N = 5) work or research in the area.

\subsection{UEQ-S findings}
\subsubsection{Overall}
\vspace{-0.4cm}
The official UEQ-S tool was used to calculate the means (\textit{M}). The values of the UEQ-S scales are derived from the arithmetic mean of the items corresponding to each dimension, whose responses are scaled from -3 to +3, and may vary within this range \cite{schrepp2023ueq}. 

Prototypes F - UX Pilot (\textit{M} = 0.720) and A - Human (\textit{M} = 0.702) have the highest average scores, indicating a more positive UX. Next, prototypes H - Human (\textit{M} = 0.601) and C - Uizard (\textit{M} = 0.569) stand out, also achieving favorable evaluations, although at a slightly lower level. Prototypes G - Lovable (\textit{M} = 0.353), E - Human (\textit{M} = 0.307), and B - Stitch (\textit{M} = 0.302) are in an intermediate range, suggesting a moderately positive perceived experience. These values indicate that, although they partially met user expectations, these prototypes did not achieve the same level of performance as the higher-rated prototypes. On the other hand, prototypes I - Magic Patterns (\textit{M} = 0.212), D - Human (\textit{M} = 0.121), and J - Human (\textit{M} = 0.107) have the lowest averages, approaching a neutral evaluation according to the UEQ-S interpretation. These results suggest limitations in the UX of prototypes, possibly related to pragmatic or hedonic aspects, or both.

\subsubsection{Analysis per UEQ-S item}

For the analysis of each item per prototype, we also used the UEQ-S data analysis tool. The overall results are shown in the Table \ref{tab:itens}.

\begin{table}[ht]
\footnotesize
\centering
\caption{Average of each item of the UEQ-S per prototype (A...J). Items 1--4 correspond to Pragmatic Quality, while items 5--8 to Hedonic Quality.}
\label{tab:itens}
\resizebox{\textwidth}{!}{%
\begin{tabular}{|p{3cm}|c|c|c|c|c|c|c|c|c|c|}
\hline
\textbf{Item} & \textbf{A} & \textbf{B} & \textbf{C} & \textbf{D} & \textbf{E} & \textbf{F} & \textbf{G} & \textbf{H} & \textbf{I} & \textbf{J} \\ \hline
1. Obstructive / Supportive & +1.5 & +1.1 & +0.9 & +0.9 & +1.1 & +1.3 & +1.2 & +1.1 & +0.8 & +0.8\\ \hline
2. Complicated / Easy & +1.7 & +1.5 & +0.5 & +1.0 & +1.4 & +1.0 & +1.2 & +0.9 & +0.8 & +0.6\\ \hline
3. Inefficient / Efficient & +1.7 & +1.3 & +1.1 & +0.9 & +1.3 & +1.3 & +1.3 & +0.9 & +0.6 & +0.7\\ \hline
4. Confusing / Clear & +1.6 & +1.2 & +0.7 & +0.8 & +1.3 & +0.9 & +1.1 & +0.8 & +0.7 & +0.6\\ \hline
5. Boring / Exciting & +0.1 & -0.5 & +0.4 & -0.7 & -0.4 & +0.6 & -0.4 & +0.4 & -0.2 & -0.3\\ \hline
6. Not interesting / Interesting & +0.6 & -0.1 & +0.8 & -0.6 & -0.4 & +0.6 & -0.2 & +0.5 & -0.2 & -0.3\\ \hline
7. Conventional / Inventive & -0.7 & -1.0 & +0.1 & -0.7 & -0.7 & 0.0 & -0.7 & +0.2 & -0.4 & -0.7\\ \hline
8. Usual / Leading edge & -0.8 & -1.1 & 0.0 & -0.6 & -1.1 & 0.0 & -0.8 & -0.1 & -0.4 & -0.5\\ \hline
\end{tabular}%
}
\end{table}

The analysis of the ten prototypes reveals positive evaluations in pragmatic UX aspects (items 1–4) and neutral or negative assessments in hedonic aspects (items 5–8), as indicated in Table \ref{tab:itens}. Thus, the results show that the interfaces were judged usable, functional, and understandable, but had limitations in arousing emotional engagement, interest, and in conveying aesthetic and conceptual innovation.

Regarding the pragmatic aspects, we observed that all prototypes were perceived as conduits (averages above +0.8), indicating fluid navigation flows, with prototypes F and G standing out. Item 2 showed positive results in most prototypes, particularly prototypes A and B, indicating and reinforcing a perception of high usability, intuitiveness, and ease of use. Item 3 maintained the same pattern, indicating efficiency in task execution and the correct arrangement of interface elements, especially in prototypes A, B, E, F, and G, which averaged 1.3 or higher. Item 4 showed results between +0.6 and +1.2 for most prototypes, indicating clear communication and good readability, though some (such as C, I, and J) need improvement.

In contrast, the hedonic items showed moderate results. Item 5 had averages near neutral (between -0.5 and +0.4), indicating the interfaces, while functional, did not evoke strong emotions or enthusiasm. Item 6 showed a mild positive trend (averages from +0.2 to +0.9), reflecting moderate interest. Items 7 and 8 mostly yielded neutral or negative averages (from -1.1 to +0.4), suggesting limited perceived originality and innovation, especially for prototypes B and E, which, though functional, require greater originality and design.

Thus, prototypes A, B, and F overall achieved the best performance due to their clarity, ease of use, and efficiency, thereby ranking high in pragmatic quality. Prototypes C, D, E, G, and H obtained intermediate results, garnering positive but unremarkable assessments, reflecting solid yet unengaging interfaces. Finally, prototypes I and J received predominantly neutral feedback, indicating acceptable usability but minimal innovation or stimulation compared to the others.

\vspace{-0.4cm}
\subsubsection{Analysis by prototype}

\begin{figure}[ht]
    \centering
    \caption{Comparison of UEQ-S results across prototypes A--J, showing mean scores for Pragmatic Quality, Hedonic Quality, and Overall UX.}    
    \label{fig:praghed}
    \fbox{\includegraphics[width=0.8\textwidth]{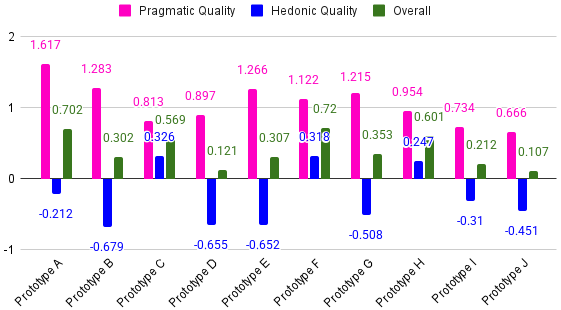}}\\
\end{figure}

For the prototype analysis, we calculated the average of the pragmatic and hedonic quality items and compared it with the overall average results. The results are shown in Figure \ref{fig:praghed}. The figure presents, on the X-axis, the ten prototypes evaluated (A to J), while the Y-axis represents the average values obtained on the UEQ-S scales, which range approximately from -1 to +2, where positive values indicate more favorable evaluations and negative values indicate less positive perceptions of UX. Each prototype has three bars: pragmatic quality (pink), hedonic quality (blue), and overall (green). 

The results show a consistent pattern of positive or neutral pragmatic evaluations, tending towards the positive, across all prototypes, with values ranging approximately from 0.666 to 1.617, indicating that, regardless of the visual design adopted, participants perceived the prototypes as clear, understandable, and functional. Prototype A (Human) presented the highest pragmatic average (\textit{A}=1.617), followed by Prototypes B (Stitch), E (Human), and G (Lovable), which also achieved values above 1.2, suggesting a strong perception of efficiency and organization.

In contrast, hedonic quality predominantly showed neutral values, tending towards negative values, indicating a lower perception of attractiveness or originality. Most prototypes obtained averages between -0.212 and -0.679, as observed in Prototypes A, B, D, E, G, I, and J. Only three prototypes stood out positively in this dimension: Prototype C - Uizard (\textit{A}=0.326), Prototype F - UX Pilot (\textit{A}=0.318), and, to a lesser extent, Prototype H - Human (\textit{A}=0.247), which presented neutral hedonic averages tending towards positive, suggesting a greater perception of originality or visual appeal.

The overall evaluation partially follows this distribution. Prototype F presented the highest overall average (0.720), followed by Prototype A (0.702) and Prototype H (0.601). These results indicate that prototypes combining good pragmatic quality with less negative or slightly positive hedonic evaluations tended to achieve higher overall scores. However, prototypes with strongly negative hedonic quality, despite a high pragmatic quality, such as B, E, and G, presented lower overall averages.

Thus, Figure \ref{fig:praghed} shows a clear dissociation between pragmatic and hedonic qualities. While all prototypes demonstrated consistent usability, only a few balanced functionality with emotional and aesthetic attributes. This result reinforces the idea that interfaces perceived as efficient are not necessarily seen as engaging or innovative, underscoring the importance of considering both functional and experiential aspects in interface design.

\subsection{Friedman and Durbin-Conover test}

The Friedman test indicated a statistically significant difference between the prototypes ($\chi^2 = 47.8$, gl = 9, $p < 0.001$), indicating that they were not perceived as equivalent. However, the test only identifies overall differences among conditions and does not indicate which specific groups differ from each other, requiring post hoc multiple-comparison tests to identify them. The highest averages were for prototypes F, A, and H, while D, I, and J presented the lowest averages and medians, as shown in the Table \ref{tab:estatistica_descritiva}.

\begin{table}[ht]
\centering
\footnotesize
\caption{Descriptive statistics of the overall scores for prototypes A--J after the Friedman test analysis.}
\label{tab:estatistica_descritiva}
\begin{tabular}{lcc}
\hline
\textbf{Prototype} & \textbf{Average (M)} & \textbf{Median (Md)} \\
\hline
Prototype A & 0.704 & 0.630 \\
Prototype B & 0.302 & 0.380 \\
Prototype C & 0.571 & 0.630 \\
Prototype D & 0.121 & 0.130 \\
Prototype E & 0.307 & 0.190 \\
Prototype F & 0.721 & 0.940 \\
Prototype G & 0.354 & 0.250 \\
Prototype H & 0.602 & 0.565 \\
Prototype I & 0.212 & 0.130 \\
Prototype J & 0.107 & 0.000 \\
\hline
\end{tabular}
\end{table}

Table \ref{tab:durbin_conover} shows a pairwise comparison matrix between the prototypes based on the Friedman test followed by the Durbin-Conover post-test. The Friedman test previously indicated statistically significant differences among the prototypes evaluated, demonstrating that not all were perceived the same way by the participants. However, this test does not allow identifying between which pairs of prototypes these differences occur. For this reason, the Durbin-Conover post hoc test was applied, which performs multiple comparisons between all pairs of prototypes. The values presented in the table correspond to the p-values of the pairwise comparisons. The significance level was 5\% ($p < 0{.}05$). Thus, values less than 0.05 indicate a statistically significant difference between the compared prototypes, while larger values suggest that there is insufficient evidence to affirm that the prototypes were perceived differently. To facilitate reading, significant values are highlighted in bold, and the symbol “--” indicates a comparison of the prototype with itself.

\begin{table}[ht]
\centering
\caption{Paired comparison matrix between prototypes based on the Friedman test followed by the Durbin-Conover post-test.}
\label{tab:durbin_conover}
\resizebox{\textwidth}{!}{
\begin{tabular}{ccccccccccc}
\hline
\textbf{Prototype} & \textbf{A} & \textbf{B} & \textbf{C} & \textbf{D} & \textbf{E} & \textbf{F} & \textbf{G} & \textbf{H} & \textbf{I} & \textbf{J} \\
\hline
\textbf{A} & -- & \textbf{0.003} & 0.226 & \textbf{$<\!0.001$} & \textbf{$<\!0.001$} & 0.890 & \textbf{0.015} & 0.465 & \textbf{0.001} & \textbf{$<\!0.001$} \\
\textbf{B} & \textbf{0.003} & -- & 0.084 & 0.338 & 0.641 & \textbf{0.005} & 0.614 & \textbf{0.028} & 0.762 & 0.059 \\
\textbf{C} & 0.226 & 0.084 & -- & \textbf{0.007} & \textbf{0.028} & 0.284 & 0.221 & 0.632 & \textbf{0.043} & \textbf{$<\!0.001$} \\
\textbf{D} & \textbf{$<\!0.001$} & 0.338 & \textbf{0.007} & -- & 0.623 & \textbf{$<\!0.001$} & 0.144 & \textbf{0.002} & 0.512 & 0.351 \\
\textbf{E} & \textbf{$<\!0.001$} & 0.641 & \textbf{0.028} & 0.623 & -- & \textbf{0.001} & 0.332 & \textbf{0.008} & 0.870 & 0.154 \\
\textbf{F} & 0.890 & \textbf{0.005} & 0.284 & \textbf{$<\!0.001$} & \textbf{0.001} & -- & \textbf{0.022} & 0.553 & \textbf{0.002} & \textbf{$<\!0.001$} \\
\textbf{G} & \textbf{0.015} & 0.614 & 0.221 & 0.144 & 0.332 & \textbf{0.022} & -- & 0.089 & 0.420 & \textbf{0.017} \\
\textbf{H} & 0.465 & \textbf{0.028} & 0.632 & \textbf{0.002} & \textbf{0.008} & 0.553 & 0.089 & -- & \textbf{0.012} & \textbf{$<\!0.001$} \\
\textbf{I} & \textbf{0.001} & 0.762 & \textbf{0.043} & 0.512 & 0.870 & \textbf{0.002} & 0.420 & \textbf{0.012} & -- & 0.112 \\
\textbf{J} & \textbf{$<\!0.001$} & 0.059 & \textbf{$<\!0.001$} & 0.351 & 0.154 & \textbf{$<\!0.001$} & \textbf{0.017} & \textbf{$<\!0.001$} & 0.112 & -- \\
\hline
\end{tabular}
}
\end{table}

In general, Prototype F showed significantly superior performance compared to several other prototypes, including B, D, E, G, I, and J. Prototype A differed significantly from D, E, G, I, and J, indicating more favorable evaluations relative to these prototypes. Prototype J, which had the lowest mean and median in the descriptive analyses, differed significantly from several prototypes with better performance, including A, C, F, and H. Significant differences were also observed between intermediate prototypes, such as C–D, C–E, D–F, E–F, and H–I, showing that the participants' perceptions were not limited only to the contrast between the prototypes with the best and worst performance, but also involved variations between average levels of evaluation. However, some comparisons did not show statistically significant differences, such as between A–C, B–C, D–E, and I–J, suggesting similar perceptions among these pairs of prototypes. The results indicate that the outstanding performance of prototypes F, A, and H contrasts with the lower ratings for D, I, and J, suggesting differences in the experience provided by each prototype.

\section{Discussion}

The results indicate differences in the perceived UX among the evaluated prototypes, with Prototype F (UX Pilot) achieving the best overall evaluation, followed by A (Human) and H (Human), suggesting that interfaces with clearer organization and interaction flows tended to provide a more positive UX. At the same time, no consistent advantage related to authorship was observed, as both AI-generated and human-created prototypes appeared among the best and worst evaluations. This indicates that the quality of the interface was more related to the effectiveness of its structure and organization than to whether it was created by AI or by a human designer.

These findings have important implications for the use of GenAI tools in interface design. The strong performance of an AI-generated prototype suggests that such tools can produce interfaces that meet fundamental usability requirements, indicating that GenAI can be useful for rapid prototyping and generating initial interface structures, especially in early design stages where multiple alternatives need to be explored quickly. However, the results also suggest that these systems may tend to reproduce common design patterns, generating solutions that are functional but not necessarily highly innovative, reinforcing the idea that designers continue to play a crucial role in originality, creativity, and differentiation in interface design.

GenAI tools can therefore be understood as complementary resources that support the design workflow rather than replace designers. In practice, they may be useful for early-stage prototyping, generating initial layouts, and accelerating ideation processes by quickly producing functional interface proposals, reducing the time required to develop initial designs, and allowing designers to focus on refinement, contextual adaptation, and creativity. In this sense, GenAI aligns with perspectives in the literature that frame artificial intelligence as a co-creative partner in the design process \cite{10.1145/3563657.3596014,Lee_2025,10.1145/3610217}. These findings suggest that GenAI tools already demonstrate potential to support the creation of usable, structured interfaces; however, their most effective application may occur when combined with human knowledge, allowing designers to build on AI-generated ideas while introducing creativity and innovation that enhance the overall UX.

\section{Final Consideration}
\label{sec:cons}

This study investigated the UX of prototypes created using GenAI tools. To this end, we conducted a survey to assess experience using the UEQ-S. The results demonstrate that, although there are statistically significant differences between the prototypes, authorship, whether human or AI, did not prove to be a determining factor in the quality of the perceived experience. Both AI-generated prototypes and those created or modified by humans showed satisfactory performance in pragmatic aspects and limitations in hedonic aspects. Thus, our study showed that differences in UX stem more from the specific characteristics of each solution than from its authorship, thereby contributing to understanding the perceptual impacts of GenAI on interface prototyping. 

Regarding limitations, the main ones were the character limit for inserting prompts in the tools, which required refinement to make them more direct. There were also limitations regarding desktop prototyping support in tools and restrictions on functionality for non-paying users. Another limitation concerns the evaluators, who were undergraduate students; although they are familiar with digital systems, their technical background may influence how interface qualities are perceived, limiting the generalizability of the findings to other user groups. Besides, exposing participants to multiple similar artifacts may have induced issues such as loss of attention and comparison effects when viewing similar artifacts. Also, the prototypes evaluated were static, meaning that participants could only observe the interface screens without interacting with them. This characteristic may have influenced perceptions of the experience's hedonic aspects, as qualities related to stimulation and engagement are often more evident during interactive use. Furthermore, all five human-created prototypes were developed by a single researcher.

The next steps in the research include conducting a qualitative analysis to identify recurring patterns and highlight visual, structural, and conceptual elements that influence perceptions of authorship and the UX. Our goal is to understand which characteristics are associated with human production and which are attributed to AI-generated production.

\bibliographystyle{sbc}
\bibliography{sbc-template}

\end{document}